\documentclass[aps,preprint,tightenlines,nofootinbib]{revtex4}

\usepackage{epsfig}

\newcommand{\lsim}{\raisebox{-0.13cm}{~\shortstack{$<$ \\[-0.07cm] $\sim$}}~}
\newcommand{\gsim}{\raisebox{-0.13cm}{~\shortstack{$>$ \\[-0.07cm] $\sim$}}~}

\begin{document}

\preprint{Edinburgh 2003/13, MPP-2003-49, PSI-PR-03-14}

\title{Higgs Radiation off Bottom Quarks at the Tevatron and the LHC}

\author{Stefan Dittmaier}
\affiliation{Max-Planck-Institut f\"ur Physik (Werner-Heisenberg-Institut), 
F\"ohringer~Ring~6, D-80805~M\"unchen, Germany}

\author{Michael Kr\"amer}
\affiliation{School of Physics, The University of Edinburgh,
Edinburgh EH9 3JZ, Scotland}

\author{Michael Spira}
\affiliation{Paul Scherrer Institut PSI, CH-5232 Villigen PSI, Switzerland}

%\date{\today}

\begin{abstract} 
Higgs-boson production in association with bottom quarks, $\,p\bar p/pp
\to b \bar bH+X,$ is one of the most important discovery channels for
supersymmetric Higgs particles at the Tevatron and the LHC. We have
calculated the next-to-leading order QCD corrections to the parton
processes $q\bar q,gg \to b\bar bH$ and present results for total
cross sections and for distributions in the transverse momenta of the
bottom quarks. The QCD corrections reduce the renormalization and
factorization scale dependence and thus stabilize the theoretical
predictions, especially when the Higgs boson is produced in
association with high-$p_T$ bottom quarks. The next-to-leading order
predictions for the total cross section are in reasonable numerical
agreement with calculations based on bottom-quark fusion $b \bar b \to
H$.
\end{abstract}

%\pacs{Valid PACS appear here}

\maketitle

\noindent
The Higgs mechanism~\cite{Higgs:1964ia} is a cornerstone of the
Standard Model (SM) and its supersymmetric extensions. The masses of
the fundamental particles, electroweak gauge bosons, leptons, and
quarks, are generated by interactions with Higgs fields. The search
for Higgs bosons is thus one of the most important endeavours in
high-energy physics and is being pursued at the upgraded
proton--antiproton collider Tevatron~\cite{Carena:2000yx} with a
centre-of-mass (CM) energy of $1.96$~TeV, followed in the near future
by the proton--proton collider LHC~\cite{atlas_cms_tdrs} with $14$~TeV
CM energy.

Various channels can be exploited to search for Higgs bosons at hadron
colliders. Higgs radiation off bottom quarks~\cite{Raitio:1978pt}
\begin{equation}
p\bar p / pp \to b\bar bH\;+\;X \qquad \mbox{via} \qquad
q\bar q,gg \to b\bar bH
\label{eq:procs}
\end{equation}
is the dominant Higgs-boson production mechanism in supersymmetric
theories at large $\tan\beta$, where the bottom--Higgs Yukawa coupling
is strongly enhanced. The parameter $\tan\beta = v_2/v_1$ is the ratio
of the vacuum expectation values of the two Higgs fields generating
the masses of up- and down-type particles in supersymmetric extensions
of the SM. $H = H_{\rm SM}, h^{0}, H^{0}$ may denote the SM Higgs
boson or any of the CP-even neutral Higgs bosons of supersymmetric
theories. In the SM the Yukawa coupling strength is given by
$g_{f\!f\!H,\mathrm{SM}}=m_f/v$, where $m_f$ denotes the mass of
fermion $f$ and $v\approx246~\mathrm{GeV}$ is the vacuum expectation
value of the SM Higgs field. In the minimal supersymmetric extension
of the SM (MSSM) the Yukawa couplings are modified by the factors
$\hat g_{f\!fH}=g_{f\!f\!H,\mathrm{MSSM}}/g_{f\!f\!H,\mathrm{SM}}$,
\begin{equation}
\displaystyle
\hat g_{uuh^0} =  \frac{\cos\alpha}{\sin\beta}, \quad
\hat g_{uuH^0} =  \frac{\sin\alpha}{\sin\beta}, \quad
\hat g_{ddh^0} = -\frac{\sin\alpha}{\cos\beta}, \quad
\hat g_{ddH^0} =  \frac{\cos\alpha}{\cos\beta},
\end{equation}
where $v=\sqrt{v_1^2+v_2^2}$ and $\alpha$ is the mixing angle of the 
CP-even Higgs fields $h^0$ and $H^0$~\cite{Gunion:1984yn}. 

Leading-order (LO) predictions for the cross sections (\ref{eq:procs})
are plagued by considerable uncertainties due to the strong dependence
on the renormalization and factorization scales, introduced by the QCD
coupling, the parton densities, and the running $b$-quark mass in the
bottom--Higgs Yukawa coupling. In this letter we present the first
calculation of the next-to-leading order (NLO) QCD corrections to
associated $b\bar bH$ production at hadron colliders through the
parton processes $q\bar q,gg \to b\bar bH$. 

Note that the inclusive cross-section prediction for the process
$gg\to b\bar bH$ develops potentially large logarithmic terms $\propto
\ln(m_b/Q)$. The logarithms arise from the splitting of gluons into 
$b\bar b$ pairs where the $b$-quark mass acts as a cutoff for the
collinear singularity. The large scale $Q$ corresponds to the upper
limit of the collinear region up to which factorization is valid. It
has been argued that $Q$ is of the order $M_H/4$ or
less~\cite{Rainwater:2002hm}. The $\ln(m_b/Q)$ terms can be summed to
all orders in perturbation theory by introducing bottom parton
densities~\cite{Barnett:1987jw}, provided that the $b$-quarks are
produced predominantly at small transverse momentum. In this
approach~\cite{Dicus:1998hs,Campbell:2002zm,Harlander:2003ai}, the
counting of perturbative orders involves both the strong coupling
$\alpha_{\mathrm{s}}$ and the logarithm $\ln(m_b/Q)$. The LO process
is Higgs-boson production through $b\bar b$ fusion, $b\bar b\to
H$. The first order corrections comprise the ${\cal O}(\alpha_{\rm
s})$ corrections to $b\bar b\to H$ and the LO process $gb\to bH$,
which is suppressed with respect to $b\bar b\to H$ by
$1/\ln(m_b/Q)$. The second-order corrections, involving $b\bar b\to H$
at NNLO, have been completed recently \cite{Harlander:2003ai} and
strongly reduce the spurious scale dependence of the total
cross-section prediction.

Higgs searches in the $b\bar{b}H$ channel, however, do in general
employ cuts on the $b$-quark transverse momenta and thus require
theoretical predictions for exclusive final states. This problem
cannot be solved within the $b$-quark density approach, since per
definition the transverse momenta of spectator $b$-quarks are
integrated out.  The solution in this case is provided by calculations
based on the parton processes $q\bar q,gg
\to b\bar bH$. In this letter we present NLO results for both
inclusive and exclusive observables, based on these parton processes.
We also include a first numerical comparison of our results with
calculations based on the $b\bar b$ fusion process.

\begin{figure}[ptb]
\epsfig{file=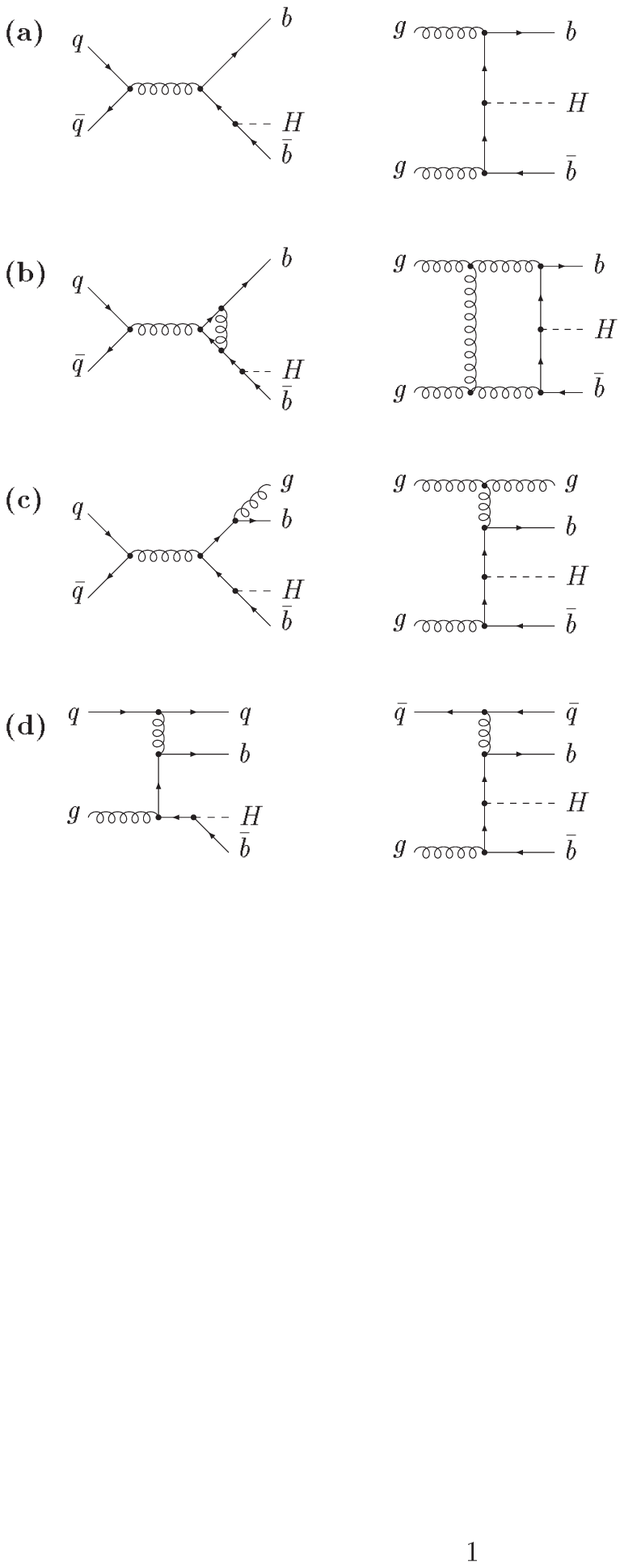,%
        bbllx=125pt,bblly=385pt,bburx=385pt,bbury=725pt,%
        scale=0.9,clip=}

\vspace*{5mm}

\caption{A generic set of diagrams (a) for the Born level, (b) for 
 virtual gluon exchange, (c) gluon radiation and (d) gluon-(anti)quark
 scattering in the subprocesses $q\bar q, gg\to b\bar bH$ etc.}
\label{fig:LOdiags}
\end{figure}

Generic diagrams that contribute to the processes (\ref{eq:procs}) at
LO are displayed in Fig.~\ref{fig:LOdiags}(a). The NLO corrections
comprise virtual one-loop diagrams, Fig.~\ref{fig:LOdiags}(b), gluon
radiation processes, Fig.~\ref{fig:LOdiags}(c), and gluon-(anti)quark
scattering reactions, Fig.~\ref{fig:LOdiags}(d). The latter two add
incoherently to the virtual corrections. The calculation of the NLO
QCD corrections to the processes $q\bar q,gg
\to Q\overline{Q}H$, where $Q$ denotes a generic heavy quark, has been
described in detail in Ref.~\cite{Beenakker:2001rj} (see also
Ref.~\cite{Dawson:2002tg}). In the following we define the physical
input parameters and quote their values as chosen for the numerical
analysis.  The renormalization of the strong coupling
$\alpha_{\mathrm{s}}(\mu)$ and the factorization of initial-state
collinear singularities are performed in the $\overline{\mathrm{MS}}$
scheme. For the calculation of the $pp$ and $p\bar p$ cross sections
we have adopted the CTEQ6L1 and CTEQ6M~\cite{Pumplin:2002vw} parton
distribution functions at LO and NLO, corresponding to
$\Lambda_5^{\mathrm{LO}} = 165$~MeV and
$\Lambda_5^{\overline{\mathrm{MS}}} = 226$~MeV at the one- and
two-loop level of the strong coupling $\alpha_{\mathrm{s}} (\mu)$,
respectively.%
\footnote{The CTEQ6 parton distribution functions involve 
five active quark flavours. In our calculation we do not include $b$
quarks as active partons. The gluon distribution is enhanced in a
scheme with four active flavours, but the effect is numerically small,
less than 5\%, and has been neglected in the present analysis.}  The
top quark is decoupled from the running of
$\alpha_{\mathrm{s}}(\mu)$. We evaluate the bottom--Higgs Yukawa
coupling with the running $b$-quark mass $\overline{m}_b(\mu)$ defined
in the $\overline{\mathrm{MS}}$ scheme in order to sum large
logarithmic corrections $\propto \ln(m_b/M_H)$ related to the
renormalization of the Yukawa coupling~\cite{Braaten:1980yq}. The
$b$-quark pole mass has been set to $m_b = 4.62$~GeV, corresponding to
a $\overline{\mathrm{MS}}$ mass $\overline{m}_b(\overline{m}_b) =
4.28$~GeV~\cite{Hagiwara:fs}. The strength of the SM Yukawa coupling
is fixed by $g_{bbH}=\overline{m}_b(\mu)/v$, where $v=
(\sqrt{2}G_F)^{-1/2}$ is the vacuum expectation value of the Higgs
field and $G_F = 1.16639\!\times\!{}10^{-5}$~GeV${}^{-2}$.

\begin{figure}[htbp]
\epsfig{file=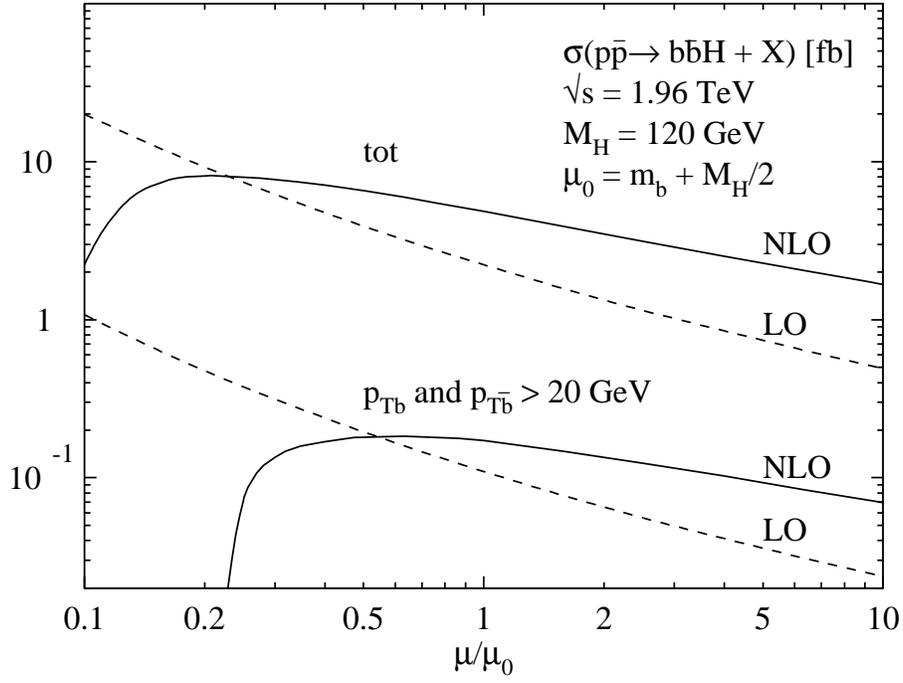,%
        bbllx=50pt,bblly=230pt,bburx=570pt,bbury=635pt,%
        scale=0.65}

\vspace*{5mm}
 
\caption{Variation of the LO and NLO cross sections with the 
 renormalization and factorization scales for $p\bar p \to b\bar bH+X$
 at the Tevatron.}
\label{fig:scale_tev}
\end{figure}

\begin{figure}[htbp]
\epsfig{file=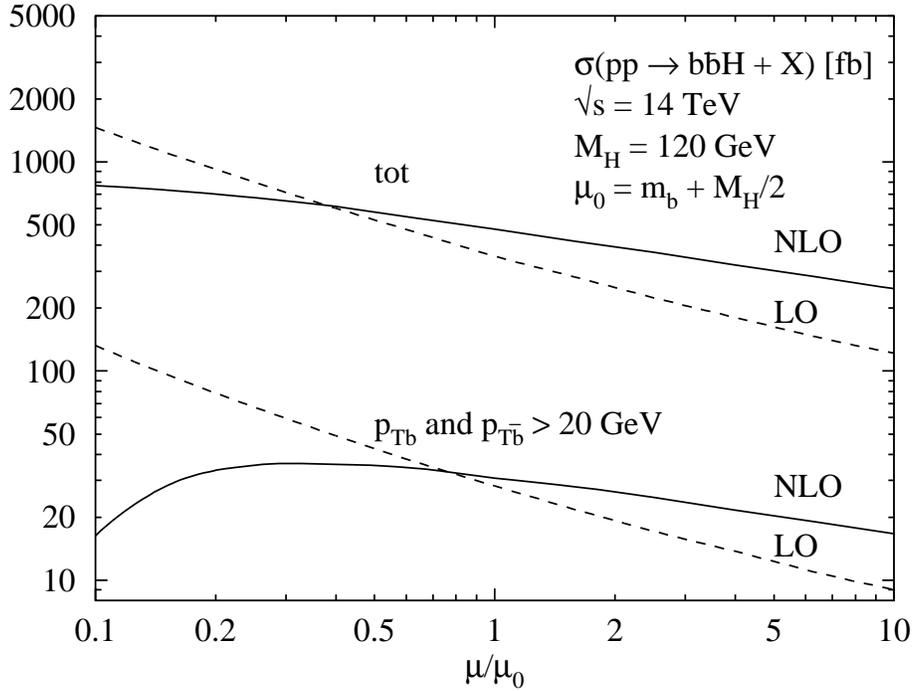,%
        bbllx=50pt,bblly=230pt,bburx=570pt,bbury=635pt,%
        scale=0.65}

\vspace*{5mm}

\caption{Variation of the LO and NLO cross sections with the 
 renormalization and factorization scales for $pp \to b\bar bH+X$ at the
 LHC.}
\label{fig:scale_lhc}
\end{figure}

In the following we present the NLO QCD predictions for associated
$b\bar bH$ production at the Tevatron and the LHC in the SM.  In LO,
the transition to the MSSM can be done by a simple rescaling of cross
sections and distributions with a factor $\hat g_{bbh^0/bbH^0}^2$,
since the LO matrix element is proportional to the bottom-quark Yukawa
coupling $g_{bbh^0/bbH^0}$. In NLO, this rescaling is spoiled by
one-loop diagrams in which the Higgs boson couples to a closed
top-quark loop (the corresponding quark loops of the first two
generations are negligible). However, our calculation reveals that
these contributions only amount to about $-5\%(-10\%)$ of the NLO
cross section at the Tevatron (LHC) in the SM. Thus, to a very good
approximation, also the cross sections and distributions including NLO
QCD corrections follow the simple LO rescaling.  We do not consider
supersymmetric-QCD and electroweak corrections in this paper.%
\footnote{For large $\tan\beta$, supersymmetric loop
corrections to the $b\bar b h^0/H^0$ vertices may become
important~\cite{Hall:1993gn}. In the present QCD analysis, the major
part of these corrections can be absorbed into effective Higgs Yukawa
couplings, since emission and reabsorption of virtual heavy
supersymmetric particles is confined to small space-time regions
compared with QCD subprocesses involving massless gluons.}  For a
consistent comparison of LO and NLO results, all quantities
($\alpha_{\rm s}$, the parton densities, the running $b$-quark mass,
and the partonic cross sections) are calculated in LO and NLO,
respectively. We first discuss the renormalization and factorization
scale dependence. All scales (including the scale of the running
$b$-quark mass) have been set equal, $\mu=\mu_F=\mu_R$, and are varied
around the central value $\mu_0 = (2m_b +
M_H)/2$. Figures~\ref{fig:scale_tev} and \ref{fig:scale_lhc} show the
scale dependence for both the total cross section and the cross
section with two high-$p_T$ bottom quarks. The reduction of the
spurious scale dependence at NLO is particularly striking for the
exclusive cross section where both $b$-quarks are required to be
produced with $p_T > 20$~GeV. A significant reduction of the scale
dependence is also observed when demanding only one of the $b$-quarks
to be produced at large $p_T$ (not shown). The residual scale
dependence of the inclusive cross section is still sizeable at NLO,
and further work is needed to improve the theoretical prediction. Note
that large logarithmic corrections may spoil the convergence of
perturbation theory if $\mu$ is chosen too low. This is evident from
the exclusive cross section at the Tevatron, Fig.~\ref{fig:scale_tev},
where the NLO prediction would even turn negative for $\mu \lsim
\mu_0/5$.  A similar observation has been made for $t\bar tH$
production at the Tevatron~\cite{Beenakker:2001rj}. We note that the
$b\bar b H$ cross section at the Tevatron and the LHC is completely
dominated by the gluon-gluon fusion and gluon-(anti)quark scattering
reactions.

\begin{figure}[htbp]
\epsfig{file=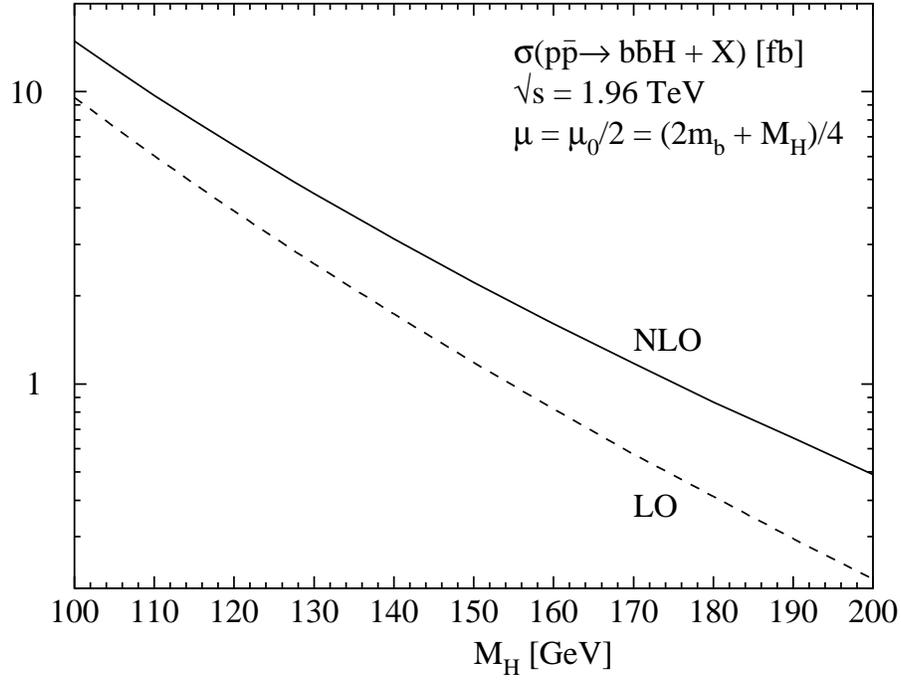,%
        bbllx=50pt,bblly=230pt,bburx=570pt,bbury=635pt,%
        scale=0.65}

\vspace*{5mm}

\epsfig{file=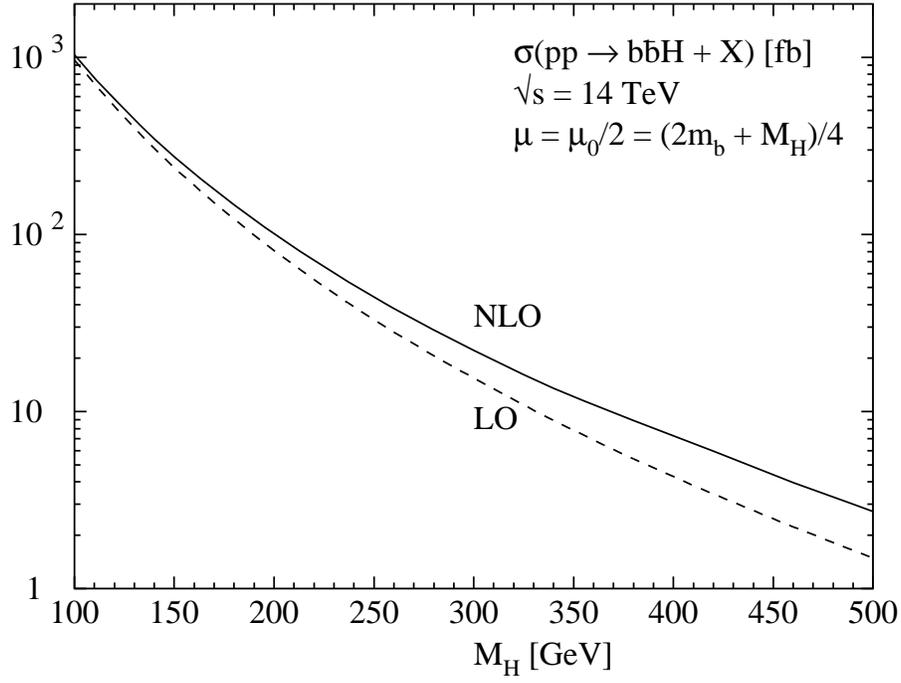,%
        bbllx=50pt,bblly=230pt,bburx=570pt,bbury=635pt,%
        scale=0.65}

\vspace*{5mm}

\caption{Total cross section for $p\bar p \to b\bar bH+X$ at the 
 Tevatron (upper figure) and $pp \to b\bar bH+X$ at the LHC (lower
 figure) as a function of the Higgs boson mass.}
\label{fig:cxn}
\end{figure}

The total cross section for associated $b\bar bH$ production at the
Tevatron and the LHC is displayed in Fig.~\ref{fig:cxn} as a function
of the Higgs-boson mass. We have set all scales equal to $\mu =
\mu_0/2 = (2m_b+M_H)/4$, which is a suitable factorization scale
choice for calculations based on the $b\bar b$ fusion
process~\cite{Dicus:1998hs,Campbell:2002zm,Harlander:2003ai}.  For
this choice of scales, the NLO QCD corrections increase the
theoretical prediction significantly, by a factor $1.6-2.3$ at the
Tevatron and $1.1-1.8$ at the LHC. Representative results for the
total cross section are listed in Table~\ref{tab:comparison} and are
compared with the NNLO calculation based on $b$-quark fusion $b\bar b
\to H$~\cite{Harlander:2003ai}.
\begin{table}
\caption{\label{tab:comparison} Total cross sections for 
$q\bar q,gg \to b\bar bH+X$ and $b\bar b \to
H+X$~\cite{Harlander:2003ai} at the Tevatron and the LHC. In the
$q\bar q,gg \to b\bar bH+X$ calculation the renormalization and
factorization scales have been set equal to $\mu = \mu_0/2 =
(2m_b+M_H)/4$ and are varied between $\mu = \mu_0/4$ (upper limit) and
$\mu = \mu_0$ (lower limit).  In the calculation based on the $b\bar b \to
H+X$ process the renormalization scale has been fixed to $\mu_R
= M_H$, while the factorization scale is varied between $0.1 M_H \le
\mu_F \le 0.7 M_H$ with $\mu_F = 0.25 M_H$ as the central scale. 
Note that the $b\bar b \to H+X$ calculation employs MRST LO and NNLO
parton distribution functions~\cite{Martin:2001es}.}
\vspace*{2mm}
\begin{tabular}{|l||c|c|c||c|c|}
\hline
 & & \multicolumn{2}{c||}{\rule[-3mm]{0mm}{8mm}
                          $\sigma(q\bar q,gg \to b\bar bH+X)$ [fb]} &
     \multicolumn{2}{c|}{\rule[-3mm]{0mm}{8mm}
                          $\sigma(b\bar b\to H+X)$ [fb]} \\ \cline{3-6} 
 & \raisebox{1.9ex}[-1.9ex]{ $M_H$~[GeV]} & 
 LO & NLO & LO & NNLO \\
 \hline\hline
         & 120 & \rule[-2.5mm]{0mm}{7mm}
         $3.9\, ^{+3.5}_{-1.7}$ &
         $6.5\, ^{+1.5}_{-1.7}$ &
         $8.6\, ^{+4.7}_{-5.0}$ & 
         $10.5 \,^{+0.3}_{-1.1}$  \\
\raisebox{1.5ex}[-1.5ex]{Tevatron} 
         & 200 & \rule[-3mm]{0mm}{7mm}
         $0.22\, ^{+0.19}_{-0.09}$ &  
         $0.49\, ^{+0.15}_{-0.15}$ &  
         $0.69\, ^{+0.20}_{-0.26}$ &  
         $0.79\, ^{+0.02}_{-0.03}$ 
         \\ \hline
         & 120 & \rule[-2.5mm]{0mm}{7mm}
         $(5.3\, ^{+2.7}_{-1.7}\,) \!\times\! 10^{2}$ & 
         $(5.8\, ^{+1.0}_{-1.0}\,) \!\times\! 10^{2}$ & 
         $(4.8\, ^{+4.3}_{-3.2}\,) \!\times\! 10^{2}$ & 
         $(7.2\, ^{+0.4}_{-1.6}\,) \!\times\! 10^{2}$ \\ 
\raisebox{1.5ex}[-1.5ex]{LHC} 
         & 400 & \rule[-3mm]{0mm}{7mm}
         $4.3\, ^{+2.4}_{-1.4}$ &   
         $7.2\, ^{+1.7}_{-1.5}$ &   
         $7.4\, ^{+2.4}_{-2.5}$ &   
         $9.8\, ^{+0.2}_{-0.4}$   \\ \hline
\end{tabular}
\end{table}
When including higher-order QCD corrections, the results obtained in
the two approaches are in reasonable numerical agreement.%
\footnote{Note that contributions involving Higgs
radiation off a closed top-quark loop have not been included in the
$b$-quark fusion calculation~\cite{Harlander:2003ai}. While these
contributions are negligible in supersymmetric theories at large
$\tan\beta$, they do affect the SM cross section. Higgs radiation of
top-quark loops has been included in the calculation presented in this
letter and reduces the cross-section prediction in the SM by
$5-10$\%.} In particular, the cross-section predictions for $b\bar b
H$ production at the LHC are compatible within their respective scale
uncertainties.  We would like to emphasize, however, that in spite of
the qualitative numerical agreement between the $q\bar q,gg \to b\bar
bH$ and $b \bar b \to H$ cross-section predictions, the intrinsic
uncertainties of both approaches have not yet been fully quantified.

\begin{figure}[htbp]
\epsfig{file=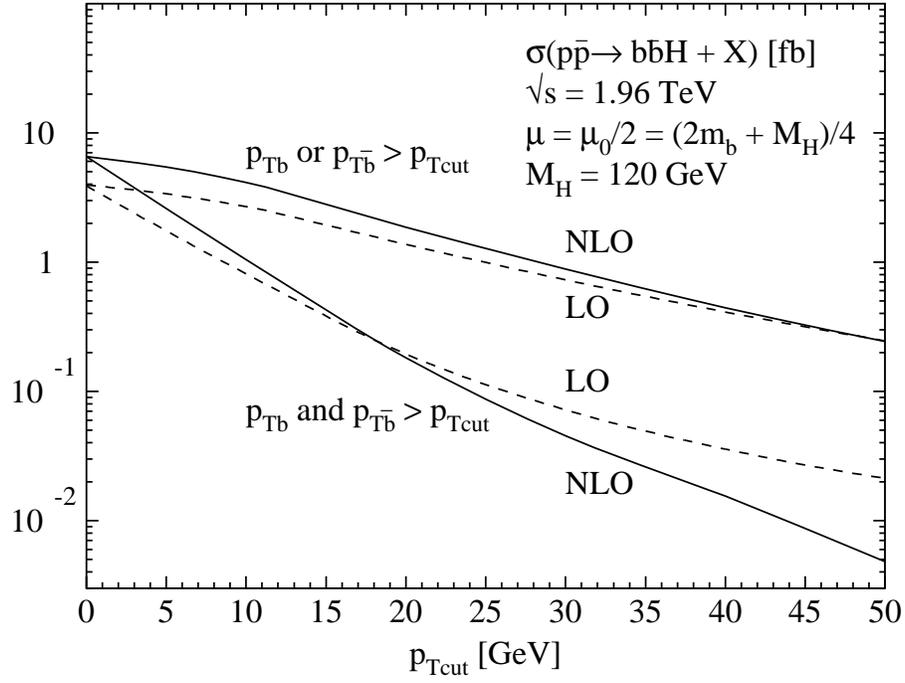,%
        bbllx=50pt,bblly=230pt,bburx=570pt,bbury=635pt,%
        scale=0.65}

\vspace*{5mm}

\caption{Cross section for $p\bar p \to b\bar bH+X$ at the Tevatron 
 with one or two high-$p_T$ bottom quarks as a function of the minimal
 $b$-quark transverse momentum.}
\label{fig:ptbtev}
\end{figure}

\begin{figure}[htbp]
\epsfig{file=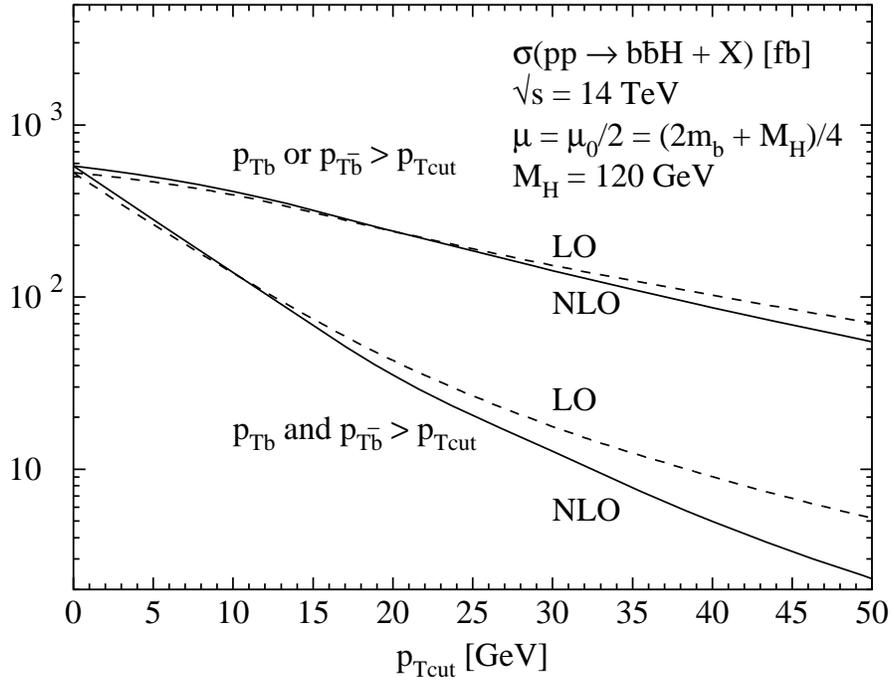,%
        bbllx=50pt,bblly=230pt,bburx=570pt,bbury=635pt,%
        scale=0.65}

\vspace*{5mm}

\caption{Cross section for $pp \to b\bar bH+X$ at the LHC with one
 or two high-$p_T$ bottom quarks as a function of the minimal
 $b$-quark transverse momentum.}
\label{fig:ptblhc}
\end{figure}

In Figures~\ref{fig:ptbtev} and \ref{fig:ptblhc} we show the cross
section with one or two high-$p_T$ bottom quarks as a function of the
minimal $b$-quark transverse momentum. We observe sizeable positive
corrections if the $b$-quarks are produced at small $p_T$. If both
$b$-quarks are required to be produced with larger $p_T
\gsim 20$~GeV the NLO corrections reduce the LO prediction.
The cross-section predictions with a single high-$p_T$ bottom quark
will be compared with calculations based on the parton process $gb\to
bH$~\cite{Campbell:2002zm} in a forthcoming publication.

In summary, the strong scale dependence of the LO predictions for the
processes $p\bar p / pp \to b\bar bH + X$, which provide important
search channels for supersymmetric Higgs bosons, requires the
inclusion of QCD corrections. For the total cross section, we find a
reduction of the spurious scale dependence at NLO, but further
improvements are needed to quantify the accuracy of the
prediction. Still, our results are in reasonable numerical agreement
with calculations based on the bottom-quark fusion process. We find a
drastic reduction of the scale dependence at NLO when the Higgs boson
is produced in association with high-$p_T$ bottom quarks. The improved
NLO predictions for the exclusive cross sections can thus be taken as
a base for experimental analyses at the Tevatron and the LHC.

\subsection*{Note added in proof}
After submission of this paper, an independent calculation of the
exclusive $b\bar{b}H$ cross section with two high-$p_T$ $b$-quarks has
been presented in Ref.~\cite{Dawson:2003kb}. The authors of
Ref.~\cite{Dawson:2003kb} have chosen a different set of input
parameters and final-state cuts, so that a direct comparison with the
results presented in our paper is not possible.  A systematic
comparison of the two calculations is, however, now in progress, and
good agreement has been found for the exclusive $b\bar{b}H$ cross
section with high-$p_T$ $b$-quarks, see Ref.~\cite{Campbell:2004pu}.
Ref.~\cite{Campbell:2004pu} also includes a comparison of the
cross-section predictions with calculations based on the parton
process $gb\to bH$.

\begin{acknowledgments}
The authors are pleased to thank W.~Beenakker, B.~Pl\"umper and
P.M.~Zerwas for their collaboration on Ref.~\cite{Beenakker:2001rj},
and W.~Beenakker and P.M.~Zerwas for comments on the manuscript. We
are grateful to the authors of Ref.~\cite{Dawson:2003kb} for a
comparison of results, and to R.~Harlander for providing us with the
results of the calculation Ref.~\cite{Harlander:2003ai}. We also thank
S.~Dawson, F.~Maltoni, T.~Plehn, L.~Reina and S.~Willenbrock for
discussions.  This work has been supported by the European Union under
contract HPRN-CT-2000-00149.
\end{acknowledgments}

\clearpage

% ---------------------------------------------------------------------
% ---------------------------------------------------------------------

% ---------------------------------------------------------------------
% ---------------------------------------------------------------------
% ---------------------------------------------------------------------
% ---------------------------------------------------------------------
% ---------------------------------------------------------------------
% ---------------------------------------------------------------------
% ---------------------------------------------------------------------
% ---------------------------------------------------------------------
% ---------------------------------------------------------------------
% ---------------------------------------------------------------------

\end{document}